\documentclass[12pt]{iopart}

\usepackage{graphicx,color}
\begin{document}

\title[Effective interactions for light nuclei]
{Effective interactions for light nuclei: an effective (field theory) approach}

\author{I. Stetcu$^1$, J. Rotureau$^2$, B.R. Barrett$^2$, and U. van Kolck$^2$}

\address{$^1$ Department of Physics, University of Washington, 
     Seattle, WA 98195}
\address{$^2$ Department of Physics, 
     University of Arizona, Tucson, AZ 85721}
\begin{abstract}
One of the central open problems in nuclear physics is the construction
of effective interactions suitable for many-body calculations.
We discuss a recently developed approach to this problem,
where one starts with an effective field theory
containing only fermion fields and formulated directly in a 
no-core shell-model 
space.
We present applications to light nuclei and to systems of a few 
atoms in a harmonic-oscillator trap.  
Future applications and extensions, as well as challenges, 
are also considered.

\end{abstract}

\maketitle

\section{Introduction}
How does the complexity of nuclear physics arise from 
the relatively simple QCD Lagrangian?
Of the open questions in nuclear structure,
this is one of the broadest and most fundamental.
The long-range goal of nuclear-structure theory is indeed to
calculate the properties of finite nuclei starting from the 
strong-interaction physics of QCD.  Although this goal is still several
years away, a confluence of nearly parallel conceptual and
computational developments is stirring unprecedented excitement
about our ability to reach it.

Tremendous progress has been made in the last decade
\cite{NVB00,NCSMrev}
in our understanding of how nuclear structure arises from the
properties of the interactions among nucleons
inside a nucleus, even though these interactions are often
modeled only in terms of {\it ad hoc} potentials.
In the previous twenty years a number of many-body techniques
has been developed for exactly solving the nuclear few-body
problem \cite{DD07,BRBetal03}.  
The application of these approaches to nuclei ranging
in mass from $A = 2$ to $A = 15$ has clearly demonstrated how
the structure of these light nuclei, {\it i.e.,} binding energies,
excitation spectra, electromagnetic moments, {\it etc.,} arise
directly from the properties of nucleon-nucleon (NN),
smaller three-nucleon (NNN),
and perhaps some tiny four-nucleon (NNNN) interactions.
One of the most tantalizing current problems is how to extend
the many-body approaches for light nuclei to medium- and heavy-mass
systems.

Independently, a framework has been developed
based on effective field theories \cite{origins,EFTrev}
to construct nuclear potentials that respect the symmetry pattern of
QCD and produce observables in a systematic and controlled
expansion in powers of momentum. These ingredients restrict the shape of
nuclear interactions, particularly in the range of pion
exchange, and encapsulate  the complicated short-range physics
into a number of ``low-energy constants''. 
Now, the first results are emerging \cite{lattQCD}, where the 
low-energy constants, so far simply fitted to data, can be 
calculated in full lattice QCD simulations.
How can one link these results to nuclear properties?

In this paper we map some of the landmarks and
crucial crossings in the long road from
QCD to nuclear structure.

\section{\bf Effective Interactions and Operators}

The general goal of microscopic nuclear-structure theory is to begin with
the free inter-nucleon interaction
and determine, using many-body quantum mechanics,
the properties of finite nuclei.  
In principle, one would like
to solve the many-body Schr\"odinger equation for all $A$ nucleons,
\begin{equation}
H |\Psi_{\alpha}\rangle  = E_{\alpha} |\Psi_{\alpha}\rangle  ,
\label{eq:fullschr}
\end{equation}
in the ``full'' Hilbert space $S$.
Here the Hamiltonian is
\begin{equation}
  H = \sum^{A}_{i=1} t_i + \sum^{A}_{i \leq j} v_{ij} 
      + \sum^{A}_{i \leq j\leq k} v_{ijk}  +\ldots, 
\label{eq:Ham2v}
\end{equation}
where the first term involves the kinetic energies $t_i$
and the subsequent terms, the potentials among an increasing number
of nucleons.
It used to be assumed that only accurate NN interactions were required,
although evidence has now accumulated for the need for NNN interactions 
\cite{Kamada01} (and perhaps NNNN interactions).
Relativistic effects do not
play a significant role in understanding low-energy nuclear structure,
and can be included as corrections around the non-relativistic limit.

In general, Eq.~(\ref{eq:fullschr}) cannot be solved in the full Hilbert
space $S$, because of the infinite number of configurations, 
so the problem must be truncated to a smaller Hilbert space $S'$ of dimension
$d$ (called the shell-model basis space or simply the model space).
Because the full space has been truncated, full space operators cannot
be used, but must be replaced with effective operators appropriate for
the given size of the model space.  In this case, $H$ must be replace
by the effective Hamiltonian $H'$, such that
\begin{equation}
H'|\Phi_{\beta}\rangle = E_{\beta}|\Phi_{\beta}\rangle ,  
\label{eq:effschr}
\end{equation}
where $|\Phi_{\beta}\rangle  = P |\Psi_{\beta}\rangle $, 
$P$ is a projection operator from
$S$ into $S'$, and the $E_{\beta}$ are a subset of dimension $d$ of the
{\it exact} eigenvalues in Eq.~(\ref{eq:fullschr}).  
The projections $\Phi_{\beta}$ are
usually not orthogonal, so one must construct the biorthogonals
$\tilde{\Phi}_{\gamma}$, such that
\begin{equation}
\langle \tilde{\Phi}_{\gamma}|\Phi_{\beta}\rangle = \delta_{\gamma\beta}.  
\label{eq:biorth}
\end{equation}
Using the biorthogonals, one can easily obtain the $H'$ that satisfies
Eq.~(\ref{eq:effschr}), {\it i.e.,}
\begin{equation}
H' = \sum_{\beta\epsilon{\cal S}} 
|\Phi_{\beta}\rangle E_{\beta}\langle \tilde{\Phi}_{\beta}|.
\label{eq:trunschr}
\end{equation}
It should be noted that $H'$ will usually be non-Hermitian, because of the
non-orthogonality of the $|\Phi_{\beta}\rangle$.  
All other physical operators relevant
to the nuclear system being investigated, {\it e.g.}, 
the rms radius operator, the
electromagnetic-moment operators, the transition operators, {\it etc.}, must be
renormalized in a similar manner for use in the given model space.

In heavy nuclei, a drastic projection is needed
in the standard nuclear shell model (SNSM),
consisting of an inert, closed-shell core and a few valence
nucleons.
The theoretical
construction of SNSM effective interactions and operators, both empirically and
microscopically, 
has an extremely long and large
history in nuclear physics \cite{Talmi03,BRB98}.
The microscopic approach has been more-or-less
unsuccessful due to problems connected with the
convergence of the perturbation-theory expansion
for the effective operators and because of the complexity of the calculations.

A response to these difficulties was found in the no-core shell model (NCSM)
\cite{NVB00,NCSMrev,NB96}, 
in which all $A$ nucleons in the nucleus are active.
In its usual formulation, the NCSM employs
a harmonic-oscillator (HO) single-particle basis of 
frequency $\omega$,
and nucleons are allowed to share a maximum number $N_{max}$
of oscillator quanta (the maximum principal quantum 
number of the wave functions) above the minimum-energy configurations.
It is important to include all configurations up to some total
energy.  This guarantees that all spurious center-of-mass
components will be projected from the final wave functions,
when using the Lawson procedure \cite{Lawson}.
The need for effective interactions and operators, 
in order to calculate nuclear
properties, exists whether one is performing SNSM or NCSM calculations.  
However, the physical nature of the required
effective operators will be quite different in the two cases.  
In the simpler NCSM, it involves only unitary
transformations based on Eq.~(\ref{eq:trunschr})
\cite{NB96,AFLetal08}.

Although the determination of the effective operators for NCSM calculations
is more straightforward, the generated effective
operators are still dependent upon the nature of the inter-nucleon interactions
employed in the calculations \cite{Stetcu_effOp}, 
a low-body (usually only two-body) cluster
approximation, and the size of the model space.
Thus, it is highly desirable to find a better method for calculating
these NCSM effective operators, which is not potential dependent, yet is
directly related to the underlying QCD symmetries and can always be
employed in small model spaces.

\section{Effective Field Theory and the NCSM}

The framework of effective field theory (EFT) emerged in the late 1970s 
\cite{weinberg} 
from the realization that, because in a quantum field theory
any physics problem is a many-body problem, one {\it never} has access
to the ``full'' Hilbert space $S$. It is always necessary to truncate
the Hilbert space so as to exclude states associated with energies
beyond those we can confidently access experimentally and theoretically.
This paradigm shift implies that interactions are only defined
in the context of a model space: the Platonic concept of ``the'' Hamiltonian
is not particularly useful in physics.

What is needed is a method to construct effective interactions and operators
whether or not the underlying, higher-energy physics is known.
Since relativistic quantum effects of arbitrary complexity 
exist unless they are forbidden by a symmetry,
any effective interaction or operator with assumed symmetries
should be included.
Still, the intrinsic connection between Hamiltonian and model space
would be problematic without a way to ensure that the arbitrariness
in the choice of model space does not contaminate observables. 
The projector $P$ always contains at least one dimensionful parameter,
the ultraviolet (UV) cutoff $\Lambda$,
for definiteness taken here to be a momentum.
One thus requires that the Hamiltonian $H'$ depend on $\Lambda$ in such a way
that observables at momenta $Q\ll \Lambda$ are independent of how $P$ 
is chosen,
and in particular, independent of $\Lambda$. 
This is termed renormalization-group (RG) invariance.

Nuclear physics is a great arena for these ideas, because
it is not easy to find solutions for QCD at momenta 
$Q <M_{QCD}\sim 1$ GeV. 
Yet we know the symmetries of the QCD Lagrangian.  In particular, 
chiral symmetry plays an important role, even at low energies, thanks to 
the appearance of light pions, due to spontaneous and 
small explicit breaking.
It is, thus, not difficult to write the most general Hamiltonian
with the appropriate degrees of freedom and symmetries \cite{origins},
which is a generalization to systems with more than one nucleon
of the Lagrangian used in chiral perturbation theory. 
Interactions among nucleons
consist of pion exchanges and contact interactions, which subsume
short-range dynamics (say, exchange of heavier mesons).
At very low energies even
pion exchange can be treated as short ranged, leaving only contact
interactions in the theory.

The real challenge is ordering the infinite number of interactions
so that observables can be calculated in an expansion
in powers of $Q/M_{QCD}$. A guide is provided by the RG: 
a truncation of this expansion at any given
order must respect RG invariance except for 
small errors contained in higher orders.
This approach had mostly been applied in particle physics
to systems where unitarity could be accounted for perturbatively.
In nuclear physics, the leading order (LO) must contain
non-perturbative physics to generate nuclear bound states and resonances.
Only subleading-order corrections, if truly corrections, should be treated
in perturbation theory.

Most applications of EFT in nuclear physics have been carried out
using a continuum free-particle basis \cite{EFTrev}.
This facilitates the study of reactions, in addition to structure,
but unfortunately, with foreseeable computational resources,
is limited to $A\le 4$.
To increase $A$, we need to limit the number of accessible one-body states
by introducing an additional, infrared (IR) momentum cutoff $\lambda$,
which discretizes momentum. 
This is a traditional method in QCD itself, where simulations
are carried out on lattices, and in addition to an UV cutoff
$\Lambda\sim 1/a$, with $a$ the lattice spacing, there is also
an IR cutoff $\lambda \sim 1/L$, with $L$ the lattice size.
Lattice regularization can be, and has been, applied to nuclear EFT as well 
\cite{lattEFT,Seki:2005ns,lattEFTrev},
where it is particularly suited to the study of nuclear matter
at finite temperature.

The successes of the SNSM suggest, however, that formulating the nuclear
EFT in an HO basis might be an efficient way to 
reach larger, finite nuclei. 
An UV cutoff equivalent to the cutoff used in a free-particle basis
can be defined in an HO basis: if $\mu$ is the two-body reduced mass,
$\Lambda=\sqrt{2\mu (N_{max}+3/2)\omega}$ 
is the momentum associated with the energy of the highest HO shell 
included explicitly in the two-body system in the center-of-mass frame.
The HO frequency $\omega$ defines an additional 
IR cutoff $\lambda=\sqrt{2}/b$, 
where $b=1/\sqrt{\mu\omega}$ is the HO length,
which plays a role analogous to the 
box size $L$ in a lattice discretization \cite{Seki:2005ns}. 

EFT in an HO basis is essentially
the NCSM formulated directly within the model space:
EFT provides the form of the effective interactions and operators needed
in the NCSM. 
Since NCSM is a full diagonalization approach in a basis constructed with 
HO wave functions, each 
model space is determined by two parameters:
$N_{max}$ and $\omega$, or alternatively, $\Lambda$ and $\lambda$.
Consequently, all low-energy constants 
are functions of these two parameters.
As in any EFT, they need to be determined
either from some experimental data or, eventually, from QCD itself.
Enormous progress in lattice QCD has already produced
\cite{lattQCD} NN scattering lengths, {\it albeit} at unphysical values of
the pion mass. It is reasonable to expect that the next few years 
will see scattering lengths at lower pion masses, effective ranges,
other NN quantities, and perhaps even observables involving more nucleons.
A key aspect of this work is the use of L\"uscher's formula \cite{luescher},
where the energy levels inside a box are linked to scattering
parameters.

The drawback of formulating the EFT in an HO basis is that 
the connection to scattering states becomes less obvious than
when using a continuum free-particle basis. 
As we show in the rest of this paper,
this is not an impassable roadblock, for two reasons.

First, one can always determine the low-energy constants from bound-state data
({\it e.g.,} binding energies). 
We illustrate the method in Sec. \ref{lightnuclei},
where we use data from a few bound states to fix the 
low-energy constants
and then many-body theory in the form of the NCSM to calculate
properties of larger systems \cite{us1}.
It is true, however, that
as the EFT order increases, the low-energy constants multiply,
and it is desirable to determine them using the more abundant scattering
(experimental or lattice) data.

Second, it {\it is} possible to connect the energies inside an HO well
to scattering parameters in a way \cite{busch} 
very similar to L\"uscher's formula. 
We show how this can be done in Sec. \ref{atoms} in the simpler
problem of spin-1/2 fermions without isospin \cite{us2}, which is relevant
for trapped two-state atoms close to a Feshbach resonance.

\section{Light Nuclei}
\label{lightnuclei}

The first direct application of EFT principles to the derivation of effective 
interactions in finite NCSM model spaces was presented in Ref. \cite{us1}. 
In that work, we have opted for a theory without explicit pions 
for two reasons: 
(i) for very-low-energy processes (involving momenta less that the pion mass) 
the formalism becomes very simple, and 
(ii) the same techniques would be readily applicable to the 
pionful EFT.

In pionless EFT at LO, the Hamiltonian $H'$ in Eq. (\ref{eq:effschr}) can be 
written as a sum of the relative kinetic energy, two contact interactions 
in the $^3S_1$ and $^1S_0$ NN channels, with corresponding parameters $C_0^1$ 
and $C_0^0$, and a contact three-body interaction 
in the $^2S_{1/2}$ NNN channel, with parameter $D_0$:
\begin{eqnarray}
H'&=&\frac{1}{4\mu A}\sum_{[i<j]}(\vec p_i-\vec p_j)^2
+C_0^1 \sum_{[i<j]^1}\delta(\vec r_i-\vec r_j)
+C_0^0 \sum_{[i<j]^0}\delta(\vec r_i-\vec r_j) \nonumber\\
& &
+D_0\sum_{[i<j<k]}\delta(\vec r_i-\vec r_j)\delta(\vec r_j-\vec r_k),
\end{eqnarray}
where $[i<j]$ denotes all pairs of particles,
$[i<j]^s$ pairs in the $S$-wave NN channel of spin $s$  
and $[i<j<k]$ triplets in the spin-$1/2$ $S$-wave NNN channel.
Unlike in the conventional NCSM approach based on a unitary transformation, 
in this approach the interaction in each model space has the same structure, 
\textit{i.e.}, matrix elements of the contact two- and three-body interactions.
Only the coupling constants differ in each model space; they depend
on $N_{max}$ and $\omega$:
$C_0^s(N_{max},\omega)$, $D_0(N_{max},\omega)$. The three coupling 
constants are fixed for each model space so that three observables, the 
deuteron, triton, and $^4$He binding energies, are simultaneously reproduced. 
With the Hamiltonian, thus, defined, we have investigated the energy of the 
first 
$(0^+,0)$ excited state of $^4$He as well as the $^6$Li ground-state energy.

\begin{figure}[t]
\begin{center}
\includegraphics*[scale=0.8]{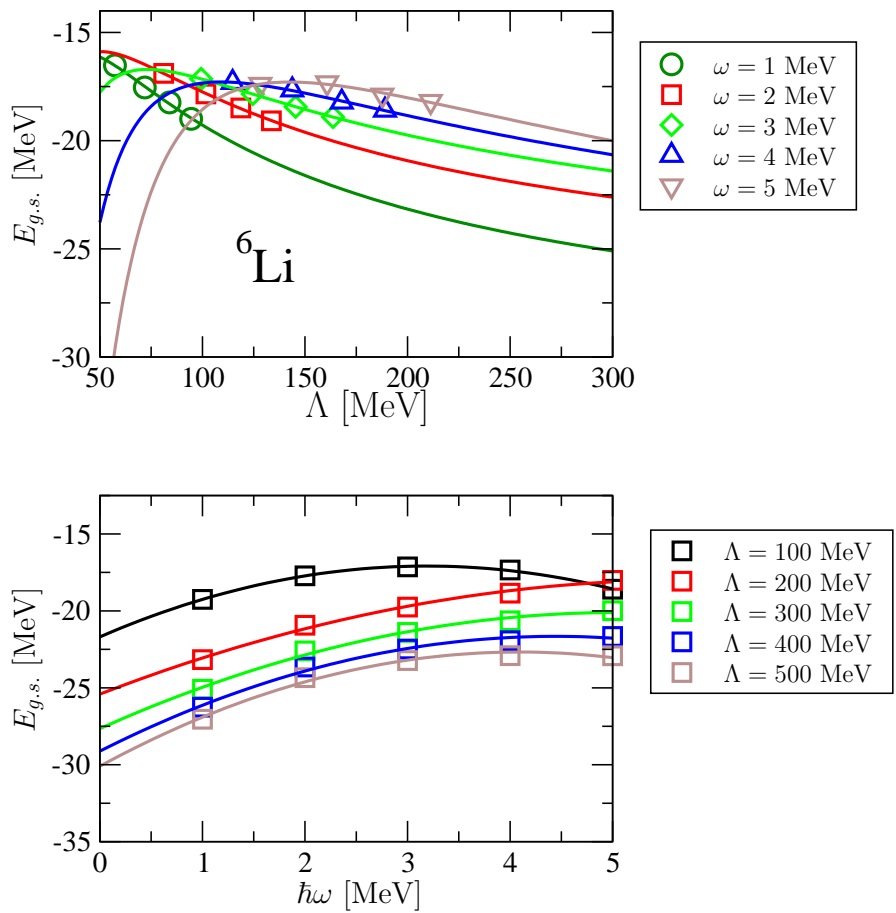}
\caption{Ground-state energy $E_{gs}$ of $^6$Li as a function of the UV cutoff
$\Lambda$ (top panel) and of the HO frequency $\omega$ (lower panel), 
all in MeV.
In the top panel we show calculated values at various frequencies
(indicated in the legend) and fits of the form discussed in the text.
In the lower panel we show values at various UV cutoffs
(indicated in the legend) and fits of a quadratic form.}
\label{fig:Li6}
\end{center}
\end{figure}

Because the 
errors associated with the terms neglected in the LO Hamiltonian $H'$ 
decrease with increasing the ultraviolet cutoff, large values of $\Lambda$ 
are desirable. While increasing $N_{max}$ makes the calculation 
impractical even for a small number of particles, increasing $\omega$ is 
not a good option either, as this would increase the errors associated 
with the infrared cutoff. Instead, for fixed $\omega$, we have calculated 
the energies of interest in the four- and six-nucleon systems for the 
maximum $N_{max}$ we were able to handle, 
extrapolating to other values of 
$\Lambda$ using $E_0(\omega)+A(\omega)/\Lambda$, an expression motivated by a 
similar running in the two-body system in the continuum \cite{EFTrev}. 
Then for fixed $\Lambda$, 
we extrapolate energies
to $\omega \to 0$, which eliminates 
infrared errors, using a quadratic formula. 
Using this approach, we have predicted that the energy 
of the first $(0^+,0)$ excited state of $^4$He is 18.5 MeV, in very good 
agreement (within 10\%) with the experimental value. (The agreement can be 
understood if we consider that this state is very close to the four-nucleon 
continuum threshold, a regime well within the limits of applicability of 
the theory.) A similar analysis for the $^6$Li ground state also produced 
results within 30\% of the experimental level, with an underbinding of about 
9 MeV. Better results can be obtained if one includes an extra term to the 
running, \textit{i.e.}, 
$E_0(\omega)+A(\omega)/\Lambda+\log(B(\omega)\Lambda)/\Lambda$. 
While the extra term essentially does not change the result in the four-body 
system, for $^6$Li the new results overbind the system by only 15\%. In 
Fig. \ref{fig:Li6}, we present the results for $^6$Li; they include the log 
term in the fit. Obviously, more investigations are necessary to pin down 
the running of the observables in the many-body system.

\section{Few-Atom Systems in Traps}
\label{atoms}

Beyond LO (or even at LO in the theory with pions \cite{NTvK}), 
the procedure 
discussed in the previous section to determine
low-energy constants becomes impractical: the number of required input 
observables increases rapidly, 
and an ever larger number of states in light nuclei is needed to adjust the 
interaction. This motivated us to devise another approach, in which the 
two-body renormalization is realized at the two-body level \cite{us2}. 
In such an approach, the low-energy scattering properties in free space 
are related to the energy spectrum of two interacting particles in a 
harmonic trap. 

The perfect testing grounds for such an approach are systems of 
two-component fermions with a large two-body scattering length $a_2$ 
in an external harmonic trap,
which can be realized experimentally with atoms
trapped by lasers in variable magnetic fields \cite{ETH}.
In this case, there is no need to take the limit $\omega \to 0$,
since $\omega$,
or equivalently the trap length $b$,
is given by the trapping laser.
In the unitary limit $b/a_2\to 0$,  
the spectrum for three particles is known \cite{werner}.

In this case, the Hamiltonian is given in LO by
\begin{eqnarray}
H'&=&\frac{\omega}{2}\left\{\sum_{i}\left[\frac{(b \vec{p}_i)^2}{2} 
+2 \left(\frac{\vec{r}_i}{b}\right)^2\right]
+2\mu b^2 C_0 \sum_{[i<j]}\delta(\vec r_i-\vec r_j)
\right\}.
\end{eqnarray}
There is just one two-body coupling constant $C_0(N_{max},\omega)$ and
three-body forces appear only at high orders.

In order to renormalize the two-body interaction in a finite model space, we 
first consider the two-body system in the center-of-mass frame,
where the relative position is denoted $\vec{r}$. 
Since the contact interaction connects $S$ states only, 
higher angular-momentum states are undisturbed from HO ones.
It is sufficient
to consider the two-body wave function 
described by a superposition of all $S$ states with the HO 
quantum number $N=2n\le N_{max}$,
\begin{equation}
\psi(\vec r)=\sum_{n=0}^{N_{max}/2}A_n\phi_n(\vec r),
\end{equation}
with $\phi_n(\vec r)$ the $S$-wave HO state with radial quantum number $n$ 
and $A_n$ a set of complex coefficients.
The $N_{max}/2+1$ unknown coefficients $A_n$ and the energy 
spectrum can be found by solving 
the Schr\" odinger equation for the relative motion,
once $C_0(N_{max},\omega)$ is determined. 

\begin{figure}[t]
\begin{center}
\includegraphics[scale=0.6]{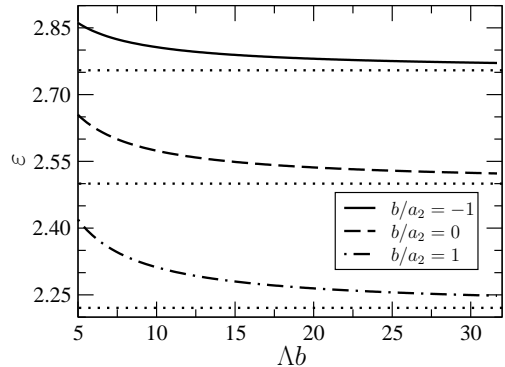}
\caption{First excited-state energy of the two-body system, $\varepsilon$,
in units of $\omega$,
as a function of the UV cutoff $\Lambda$, in units of the inverse of 
the trap length $b$.
Three interaction strengths determined by the
indicated values of $b/a_2$ are shown. 
The ground-state energy given by Eq. (\ref{busch_2b}) is used in each case 
to fix the coupling constant. 
The exact values for the first excited-state energy
are displayed with dotted lines.}
\label{fig:2b}
\end{center}
\end{figure}

The coupling $C_0(N_{max},\omega)$ can be determined by the following 
procedure. 
In order for the Schr\" odinger equation with the delta function 
to be well defined, $C_0(N_{max},\omega)$ and
the energies $\varepsilon(N_{max},\omega)$ (in units of $\omega$)
have to satisfy an
RG condition \cite{us2}, 
\begin{equation}
\frac{1}{C_0(N_{max},\omega)}=
-\frac{\mu}{\pi^{3/2}b}
\sum_{n=0}^{N_{max}/2}\frac{L_n^{(1/2)}(0)}{(2n+3/2)
-\varepsilon(N_{max},\omega)},
\label{prebusch}
\end{equation}
where $L_n^{(1/2)}(0)$ is a generalized Laguerre polynomial at the origin.
In the $\Lambda \to \infty$ limit, one can show \cite{us4} that this condition
leads to 
\begin{equation}
\frac{\Gamma(3/4-\varepsilon(\infty,\omega)/2)}
     {\Gamma(1/4-\varepsilon(\infty,\omega)/2)}
=\frac{b}{2a_2}.
\label{busch_2b}
\end{equation}
That is, the energy 
spectrum inside the trap 
depends only upon the scattering length in units of the trap length.
This formula, the HO oscillator counterpart
of the L\"uscher formula for a box with sharp boundaries,
was first derived in Ref. \cite{busch}, using a pseudopotential.
The $C_0(N_{max},\omega)$
can then be fixed so that in each model space 
one of the states obtained by the exact diagonalization
of the Schr\"odinger equation 
reproduces the corresponding value given by Eq. (\ref{busch_2b}). 
For simplicity, we match the lowest state. 
Because just one coupling constant needs to be fixed in LO, 
all the other levels can be calculated.
They satisfy Eq. (\ref{prebusch}) and
deviate from the exact value given by Eq. (\ref{busch_2b}), but the 
error decreases with increasing the size of the model space. 
As an illustration,
in Fig. \ref{fig:2b} we present 
the running of the first excited state of the two-body system as a function 
of the dimensionless 
quantity $\Lambda b$ for three selected values of the $b/a_2$ ratio. 
In all cases, the energy goes to the exact value with errors that decrease as 
$1/\Lambda b$. Faster convergence can be achieved by considering terms beyond 
leading order, that is, corrections that involve derivatives of the contact 
interaction and account for the effective range, the shape parameter, 
{\it etc.}, introduced by the model-space truncation \cite{us4}.

\begin{figure}[t]
\begin{center}
\includegraphics*[scale=0.6]{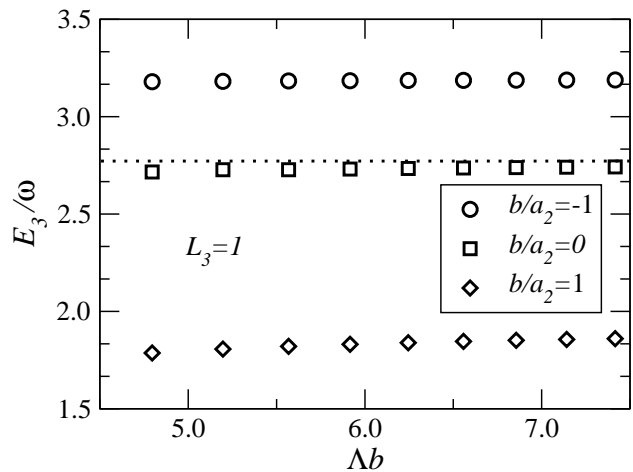}
\caption{Ground-state energy of the three-body system
in units of $\omega$
as a function of the UV cutoff $\Lambda$, in units of the inverse of 
the trap length $b$.
Three interaction strengths determined by the
indicated values of $b/a_2$ are shown. 
The two-body ground-state energy given by Eq. (\ref{busch_2b}) 
is used in each case to fix the coupling constant. 
The exact value at unitarity \cite{werner}
is displayed with a dotted line.}
\label{fig:3b}
\end{center}
\end{figure}

With the two-body interaction fixed in a truncated two-body model space, 
we turn now to the few-body problem, considering the unitary regime 
as well as a general non-vanishing $b/a_2$ value, 
for both positive and negative scattering lengths. 
The three-body solutions are obtained by a diagonalization in a finite model 
space, constructed as anti-symmetrized three-body states of HO wave functions 
(for details on the basis construction in each model space, see 
Refs. \cite{us1,us2}). 
In Fig. \ref{fig:3b}, we present the running of the energy of the
lowest three-body state (which has orbital angular momentum $L=1$),
at unitarity and at $b/a_2=\pm 1$. 
At unitarity, where a semi-analytical result exists \cite{werner}, 
our results are within a few percent of the exact results 
even for relatively modest model-space sizes. 
However, it is worth noting that the running is rather slow, having in general 
a $1/(\Lambda b)^\alpha$ dependence, with $\alpha$ a power that is state 
dependent. For the ground state at unitarity, we have shown by direct fit 
that $\alpha=1$ \cite{us2}, and in the limit $\Lambda b\to \infty$, 
we obtain the exact result \cite{werner}. 
Faster convergence, although not necessarily faster running (that is, 
same power $\alpha$, but larger coefficients in front of 
$1/(\Lambda b)^\alpha$), can be achieved if one introduces 
higher-order terms \cite{us3}.

In the non-interacting limit ($b/a_2\to-\infty$), the lowest state of 
three two-component fermions has $L=1$ (negative parity). 
As $b/a_2$ increases, this state persists as the ground state 
beyond unitarity.
However, 
around $b/a_2\approx 1.5$, the lowest state becomes an $L=0$ state \cite{us2}, 
which remains lowest as $b/a_2$ increases further. 
In the limit of large $b/a_2$, 
which approaches the untrapped case, the 
energy of the three-body ground state 
becomes $E_3\approx -1/2\mu a_2^2$. This result suggests that the system of 
three fermions is near the threshold of the $S$-wave scattering of one 
fermion on the bound state of the other two. 

Further investigation has shown that the known results in the four-body 
system \cite{alhassid}
are also reproduced, although with less accuracy because of the complexity 
of solving the four-body problem in larger model spaces. 
Nevertheless, the errors can be reduced by introducing higher-order 
corrections, and preliminary calculations show very nicely this trend 
\cite{us2,us3}.

\section{Future Applications and Challenges}
\label{challs}

As discussed in Sec. 3, our goal is to obtain effective interactions
and operators in truncated model spaces for solving the few-body 
problem (for fermions and bosons), based on the underlying assumed
QCD symmetries, thus bypassing, for example, the need for a phenomenological
potential.  As we have shown, this approach works reasonably well for 
light nuclei (Sec. 4) and even better for few-fermion systems in
an HO trap (Sec. 5).

Now that we have
achieved good results linking 
few-fermion energies in an HO trap to scattering parameters, 
we can pursue our
original idea of applying this method to nuclei.  Doing this is, 
indeed, a great challenge, as we have indicated in Secs. 3 and 4:
first, because the number of low-energy constants increases 
significantly as the EFT order increases; and, second, because
the many-body calculations ({\it i.e.,} within the NCSM) become
more and more difficult as the model-space size, as defined by
$N_{max}$, increases for larger values of $A$.
For given computational resources, it is, thus, important
to obtain better converged 
results in smaller model spaces for NCSM calculations
of heavier nuclei.

Obviously, we want to extend this approach to higher orders and include the
effective range.
In preliminary calculations for few-fermion systems in a trap 
\cite{us4,us3} we have
included corrections to the potential
up to N$^2$LO and observed an acceleration of the convergence for energies.
Corrections beyond LO are treated as perturbations and these preliminary 
results show an excellent agreement at unitarity with known results
for the three-fermion system \cite{werner} 
and with other methods for the four-fermion 
system \cite{alhassid}.
One finds that this perturbative treatment of subleading interactions,
demanded by RG consistency, gives faster convergence
than a non-perturbative treatment.

Another new idea deals with making sure that the available energy in
the many-body system is larger than the
maximum two-body energy  
employed in determining the two-body interactions.  
This is achieved by introducing a different total number of oscillator
quanta $N^{(n)}_{max}$ for each $n$-body space.
Preliminary calculations \cite{us3} for the problem of 
few-fermion systems in an HO trap indicate that one obtains 
significant gains in the converged results in smaller model 
spaces, when for a fixed value of $N^{(2)}_{max}$, 
the many-body model space is increased until convergence.
The next step is to apply this approach to few nucleon systems.

Although we have so far concentrated on binding energies, other 
bound-state observables
can be calculated with our method. Using similar techniques,
we can construct other operators
for describing observables of interest. 
It will be instructive to compare with results obtained
in the {\it traditional} NCSM \cite{Stetcu_effOp}.

Finally, an important long-term goal of this program is to eventually complete 
the link from QCD to nuclear-structure observables.
In the direction of QCD, we should
determine the
low-energy constants of the EFT expansion from lattice simulations.
In the direction of heavier nuclei, we need to understand
the limits of the pionless theory, and presumably go beyond it.
The pionless EFT is valid for low momenta ($Q< m_\pi$), and so is 
Eq. (\ref{busch_2b}). 
One of the challenges of future applications to nuclear systems is to 
extend the equation that determines the spectra of two trapped particles 
for momenta large enough to discern pions.

\section{Summary and Conclusions}

We have presented a discussion of
the first implementation of EFT principles 
directly into a many-body method, the NCSM. 
We have proposed that using EFT to construct effective 
interactions in restricted model spaces used in NCSM calculations
might provide an important step in the long road from QCD to nuclear structure.

In one implementation, we have determined the low-energy constants 
by a direct fit to ground-state binding energies in two-, three- and 
four-nucleon systems, 
and predicted an excited state in $^4$He, as well as the ground-state energy 
of $^6$Li \cite{us1}. We have presented an improved extrapolation for
the latter in Sec. \ref{lightnuclei}.
For both states, results are within the expected errors of the 
pionless EFT. Such an application becomes quickly impractical 
if one considers the subleading orders in the pionless EFT, 
or the pionful theory, as the number of low-energy constants 
that have to be determined by bound states increases considerably. 

In a second implementation, we have considered a system of trapped fermions 
\cite{us2}. Such an approach would allow the determination of 
low-energy constants in the two-body system alone, as shown in 
Sec. \ref{atoms}. We have illustrated the application of the two-body 
renormalization by computing the spectrum of three spin-1/2 fermions in a trap
at unitarity, where a semi-analytical calculation exists \cite{werner}. 
Thus, we have shown that in LO our results converge to the exact ones, 
in the limit of large UV cutoffs. 
The advantage of our approach is that it can be extended to finite two-body 
scattering lengths, the convergence of which we examined here, and 
to include other scattering parameters 
\cite{us4,us3}.

Despite the different underlying physics, the systems of trapped cold atoms 
near Feshbach resonances and of
nucleons at low energies are quite similar. One can hope that the same 
procedures can be transferred to the nuclear many-body problem,
as we discussed in Sec. \ref{challs},
thus providing a QCD-based solution 
to the open problem of constructing nuclear effective interactions.
Work in this direction
is in progress. 

\section*{Acknowledgments}
We acknowledge partial support from US
DOE grant DE-FC02-07ER41457 (IS), 
NSF grants PHY-0555396 and PHY-0854912 (BRB and JR),
and 
US DOE grant DE-FG02-04ER41338 (JR and UvK).

\end{document}